\documentclass{moriond}

\usepackage{amsmath}

\def\be{\begin{equation}}
\def\ee{\end{equation}}
\def\bea{\begin{eqnarray}}
\def\eea{\end{eqnarray}}

\newcommand{\Photo}{}

\begin{document}
\vspace*{4cm}
\title{First observation of $\Lambda_b^0 \rightarrow \Lambda K^+\pi^-$ and $\Lambda_b^0 \rightarrow \Lambda K^+K^-$ decays at LHCb}

\author{ Daniel O'Hanlon, \\ on behalf of the LHCb collaboration }

\address{Department of Physics, University of Warwick,\\
Coventry, CV4 7AL, United Kingdom}

\maketitle\abstracts{
The physics potential for $b$-baryon decays has been relatively unexplored until the advent of the LHC, and as such important questions still exist about their fundamental properties, such as whether their decays exhibit $C\!P$ violation. Presented here are observations of the decays $\Lambda_b^0 \rightarrow \Lambda K^+\pi^-$ and $\Lambda_b^0 \rightarrow \Lambda K^+K^-$, made at a significance level of 8.1 and 15.8 Gaussian standard deviations, respectively, and measurements of their branching fractions. The phase-space integrated $C\!P$ asymmetries of these decays are also measured and found to be consistent with zero. Limits are set on the branching fractions of other $\Lambda_b^0$ and $\Xi_b^0$ decays to $\Lambda h^+h^{\prime -}$ (where $h$ is a kaon or pion).}

\section{Introduction}

Despite the rapid improvement in precision measurements of $b$-baryon properties afforded by the high energy and high intensity of the Large Hadron Collider, relatively few decay modes of these particles have been exploited. In particular, of the possible decays to charmless hadronic final states, only the two body $\Lambda_b^0 \rightarrow p\pi^-$ and $\Lambda_b^0 \rightarrow pK^-$, the quasi-two-body $\Lambda_b^0 \rightarrow \Lambda \phi$, and the three-body $\Lambda_b^0 \rightarrow K^0_{\rm S}p\pi^-$ decays have been observed~\cite{Aaltonen:2008hg,LHCb-PAPER-2016-002,LHCb-PAPER-2013-061}.

The charmless decay modes are of great interest in particular, as in the Standard Model these proceed either by loop-induced amplitudes or via the Cabibbo-Kobayashi-Maskawa matrix element $V_{ub}$, and as such are expected to be suppressed. The interference between such amplitudes provides potential for large $C\!P$ violation, which in three-body decays can vary as a function of the phase-space location~\cite{LHCb-PAPER-2014-044}. More generally, decays of this type provide insights into the mechanisms of hadronisation in $b$-baryon decays.

Here we report a search for the charmless decays of $\Lambda_b^0$ and $\Xi_b^0$ baryons to $\Lambda \pi^+\pi^-$, $\Lambda K^{\pm}\pi^{\mp}$, and $\Lambda K^+K^-$ final states~\cite{Aaij:2016nrq}, using data collected by the LHCb experiment~\cite{Alves:2008zz} corresponding to $1$~$\rm{fb}^{-1}$ at a centre-of-mass energy of $7$ TeV in 2011 and $2$~$\rm{fb}^{-1}$ at a centre-of-mass energy of $8$ TeV in 2012.

\section{Analysis procedure}

The $b$-baryon candidates are reconstructed by combining two oppositely charged tracks with a $\Lambda$ candidate, where the $\Lambda$ candidates are reconstructed by combining $p$ and $\pi^-$ tracks.
To reduce backgrounds from $K_{\rm S}^0$ decays, loose particle identification (PID) criteria, based primarily on the information from the ring-imaging Cherenkov detectors, are imposed for the $p$ track. Further particle identification criteria are required for the di-hadron pair from the $\Lambda_b^0$ to separate the dataset into the different final states under study. Intermediate charm states are vetoed, however data corresponding to the $\Lambda_b^0 \rightarrow (\Lambda\pi^+)_{\Lambda_c^+}\pi^-$ decay is used as a control channel.

All signal and background yields, as well as the yield of the control mode, are obtained from a single simultaneous unbinned maximum likelihood fit to the $b$-baryon invariant mass distribution for each final state in six subsamples, corresponding to three data-taking periods and two reconstruction categories. This simultaneous fit enables information on the $K \leftrightarrow \pi$ mis-ID in each spectrum to be constrained. The regions of the $\Lambda h^+h^{\prime-}$ invariant mass spectra corresponding to the $\Lambda_b^0$ and $\Xi_b^0$ invariant masses were not inspected until the candidate selection and fit model was finalised, to avoid bias.

In addition to the mis-ID background, components are included in the fit model to account for a smoothly varying combinatorial background shape, comprised of hadrons that do not come from the $b$-baryon decay, and partially reconstructed backgrounds that peak at a lower mass than the signal, where a photon or pion is not reconstructed in the decay of a real $b$-baryon. A boosted decision tree (BDT) classifier, trained on sideband collision data and simulated signal data, is used to further suppress the combinatorial background. The input variables to this BDT were chosen such that the performance of the algorithm is optimal whilst minimising the expected discrepancy between the simulated and collision data distributions.

The efficiency of the signal selection requirements is studied using simulated data, and for the PID requirements, high-yield control samples of $D^0 \rightarrow K^-\pi^+$ and $\Lambda \rightarrow p \pi^-$ decays. In general, multi-body decays proceed via several intermediate quasi-two-body decays, as well as a non-resonant decay, and hence their phase-space distributions are non-uniform and in this case not known \emph{a priori}. The efficiency is also phase-space dependent, and as such, an efficiency correction is implemented using the product of one two-dimensional `Dalitz' efficiency histogram and three one-dimensional angular efficiency histograms. For signal modes with a significant yield, the distribution in the phase-space is obtained with the \emph{sPlot} technique~\cite{Pivk:2005}, with the $b$-baryon candidate invariant mass used as the control variable. Where no such signal is observed, the efficiency corresponding to a uniform phase-space distribution is used, and a systematic uncertainty is assigned to account for the variation across the phase-space.

\begin{figure}[h]
\centering
\includegraphics*[width=0.45\textwidth]{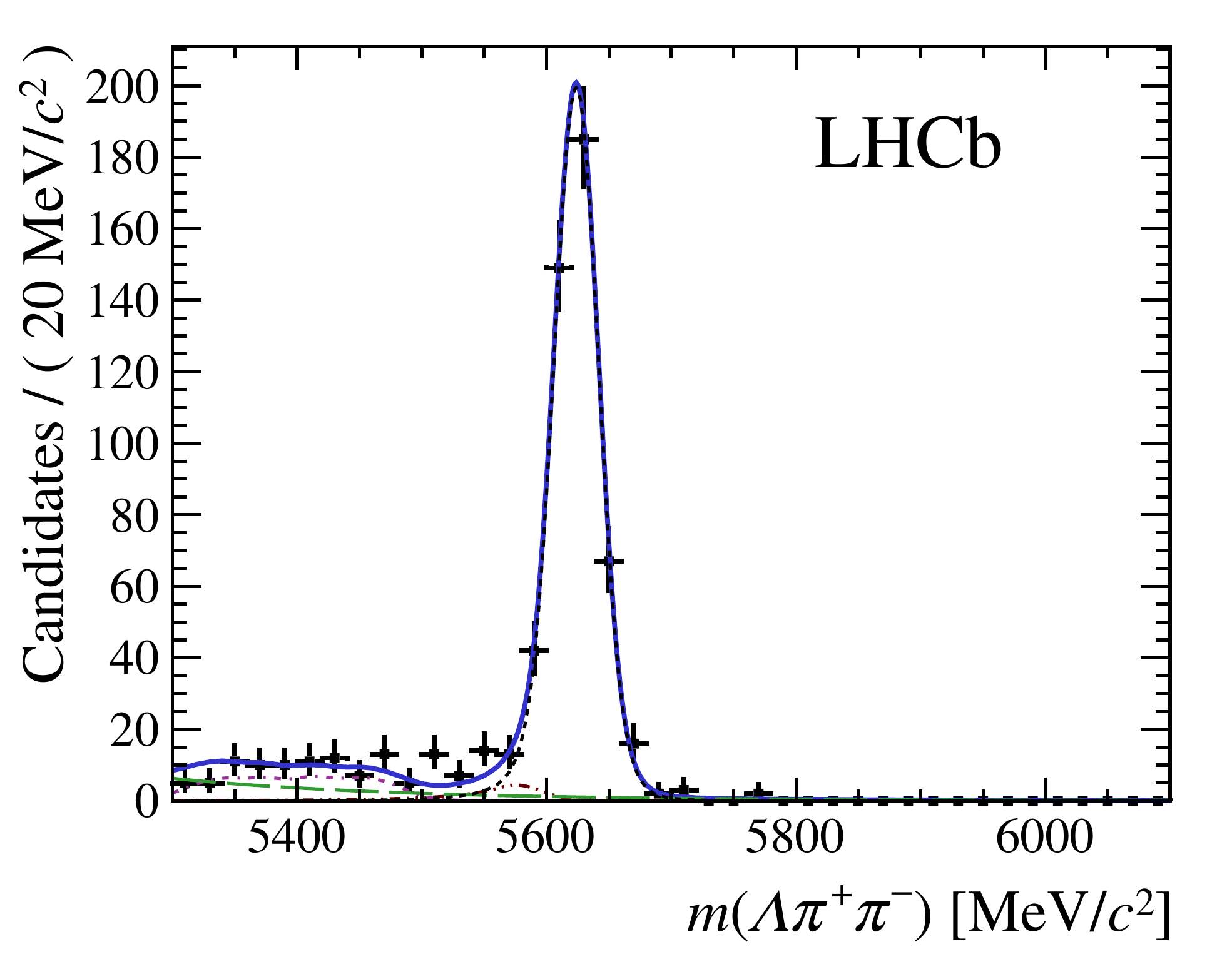}
\includegraphics*[width=0.45\textwidth]{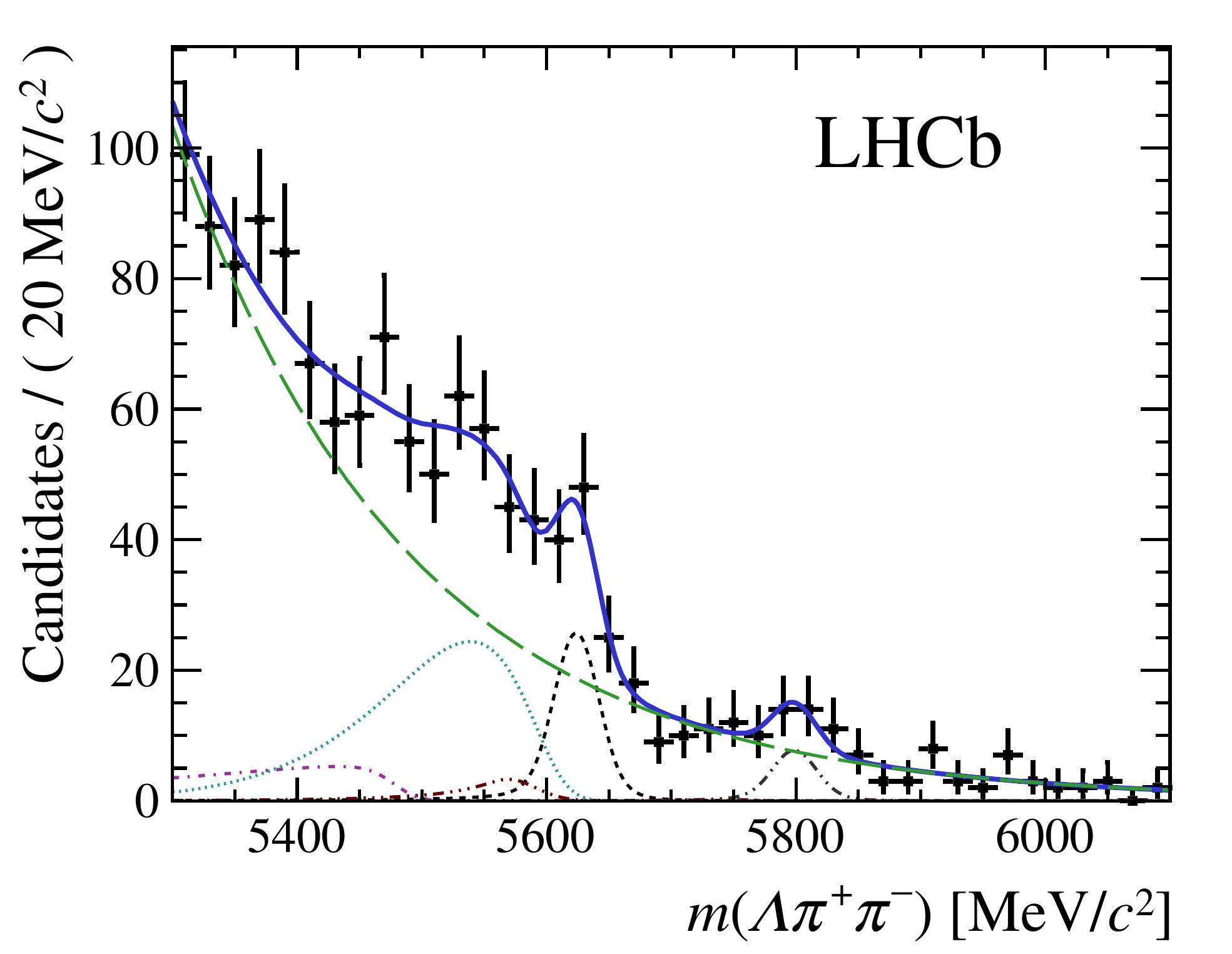}\\
\caption{\small
  Results of the fit for the $\Lambda_b^0 \to \left(\Lambda\pi^+\right)_{\Lambda_c^+}\pi^-$ control mode (left), and $\Lambda \pi^+\pi^-$ signal (right) final states, for all subsamples combined. Superimposed on the data are the total result of the fit as a solid blue line, the $\Lambda_b^0$ ($\Xi_b^0$) decay as a short-dashed black (double dot-dashed grey) line, cross-feed as triple dot-dashed brown lines, the combinatorial background as a long-dashed green line, and partially reconstructed background components with either a missing neutral pion as a dot-dashed purple line or a missing soft photon as a dotted cyan line.
}
\label{fig:dataFitspi}
\end{figure}

\section{Fit model and results}

 In the simultaneous maximum likelihood fit, the probability density function that describes each invariant mass distribution is constructed as a sum of individual components that correspond to the $\Lambda_b^0$ and $\Xi_b^0$ signal components, partially reconstructed backgrounds, the cross-feed backgrounds, and the combinatorial background.

\begin{figure}[h]
\centering
\includegraphics*[width=0.45\textwidth]{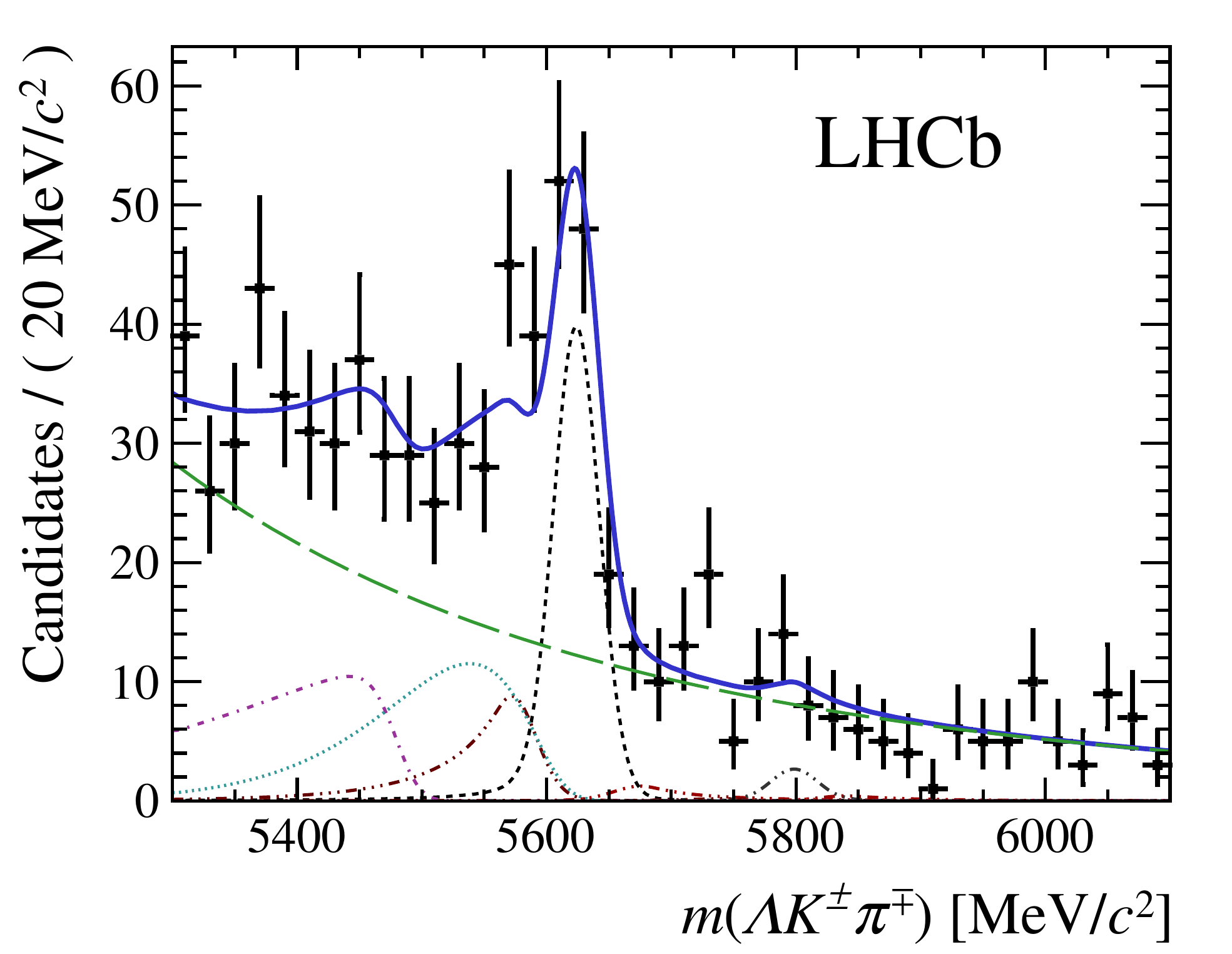}
\includegraphics*[width=0.45\textwidth]{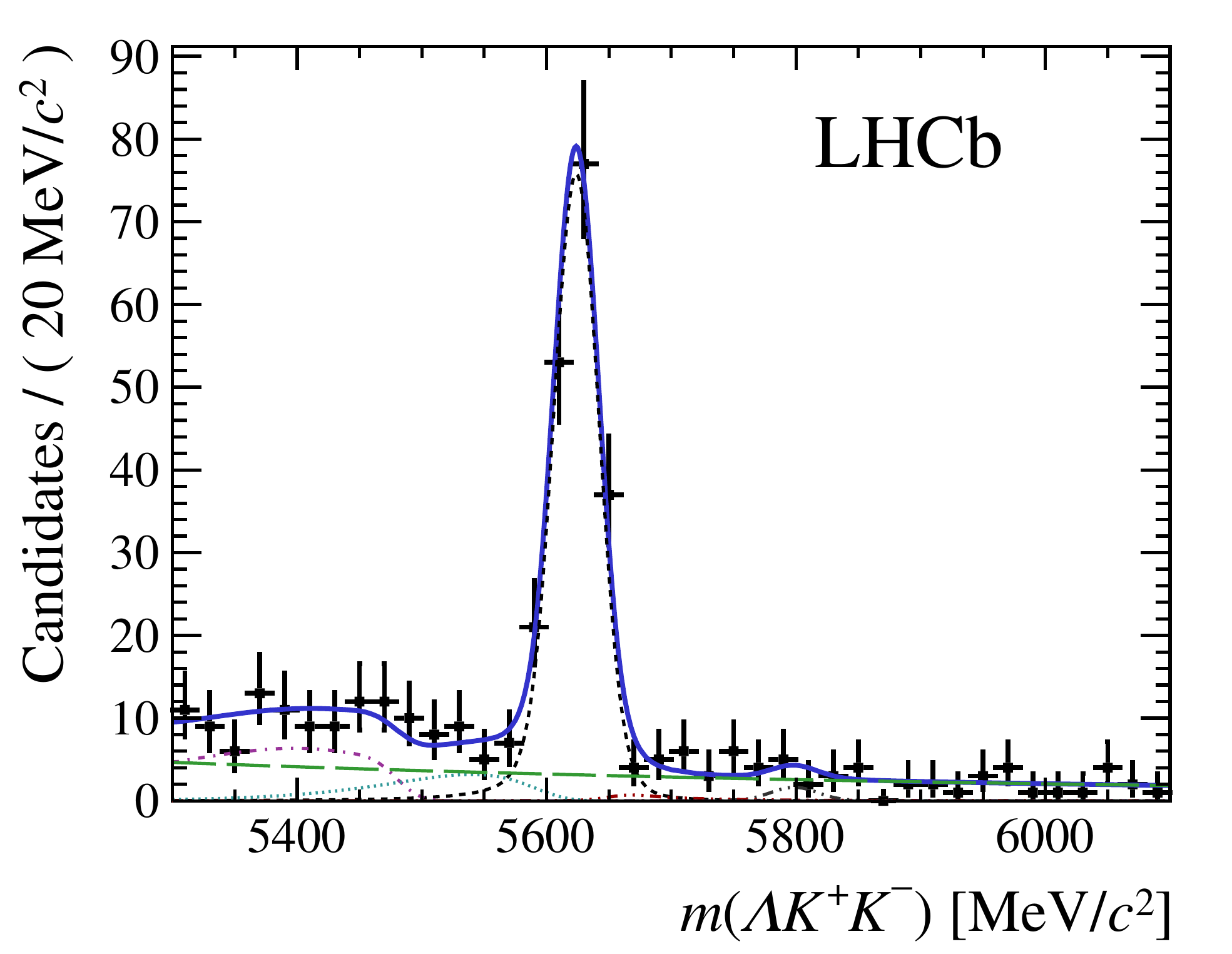}\\
\caption{\small
  Results of the fit for the $\Lambda K^{\pm}\pi^{\mp}$ signal (left), and $\Lambda K^+ K^-$ signal (right) final states, for all subsamples combined. Superimposed on the data are the total result of the fit as a solid blue line, the $\Lambda_b^0$ ($\Xi_b^0$) decay as a short-dashed black (double dot-dashed grey) line, cross-feed as triple dot-dashed brown lines, the combinatorial background as a long-dashed green line, and partially reconstructed background components with either a missing neutral pion as a dot-dashed purple line or a missing soft photon as a dotted cyan line.
}
\label{fig:dataFitsK}
\end{figure}

 The projections of this fit model on the individual invariant mass spectra for each final state can be seen in Figs.~\ref{fig:dataFitspi} and \ref{fig:dataFitsK}, where each separate component, in addition to the total sum, is indicated. When systematic uncertainties are taken into account, the statistical significance of the $\Lambda_b^0 \rightarrow \Lambda \pi^+\pi^-$ signal is $4.7\sigma$, the significance of the $\Lambda_b^0 \rightarrow \Lambda K^+\pi^-$ signal is $8.1\sigma$, the significance of the $\Lambda_b^0 \rightarrow \Lambda K^+K^-$ signal is $15.8\sigma$, and the significances of all other modes are less than $3\sigma$.

\subsection{Branching fractions}

Branching fraction central values and confidence intervals are obtained by combining the individual branching fractions for each of the six data-taking categories, taking systematic uncertainties into account. Upper limits are calculated using a Bayesian prescription, with a prior removing the unphysical region.

The results for the absolute branching fractions are
\begin{equation*}
  \begin{array}{rcr}
    {\cal B}(\Lambda_b^0 \rightarrow \Lambda \pi^+\pi^-) & = & \left(4.6 \pm 1.2 \pm 1.4 \pm 0.6 \right) \times 10^{-6}\,, \\ [0.8ex]
    {\cal B}(\Lambda_b^0 \rightarrow \Lambda K^+\pi^-) & = & \left(5.6 \pm 0.8 \pm 0.8 \pm 0.7 \right) \times 10^{-6}\,, \\  [0.8ex]
    {\cal B}(\Lambda_b^0 \rightarrow \Lambda K^+K^-) & = & \left(15.9 \pm 1.2 \pm 1.2 \pm 2.0 \right) \times 10^{-6}\,, \\
    \frac{f_{\Xi_b^0}}{f_{\Lambda_b^0}} \times {\cal B}(\Xi_b^0\rightarrow\Lambda \pi^+\pi^-) & = & \left(1.3 \pm 0.6 \pm 0.5 \pm 0.2 \right) \times 10^{-6}\,, \\
    & < & \multicolumn{1}{l}{~1.7 ~(2.1) \times 10^{-6} \ \rm{at \ 90~(95)\,\% \ confidence \ level} \, ,} \\
    \frac{f_{\Xi_b^0}}{f_{\Lambda_b^0}} \times {\cal B}(\Xi_b^0\rightarrow\Lambda K^-\pi^+) & = & \left({-}0.6 \pm 0.5 \pm 0.3 \pm 0.1 \right) \times 10^{-6}\,, \\
    & < & \multicolumn{1}{l}{~0.8 ~(1.0) \times 10^{-6} \ \rm{at \ 90~(95)\,\% \ confidence \ level} \, ,} \\
    \frac{f_{\Xi_b^0}}{f_{\Lambda_b^0}} \times {\cal B}(\Xi_b^0\rightarrow\Lambda K^+K^-) & < & \multicolumn{1}{l}{~0.3 ~(0.4) \times 10^{-6} \ \rm{at \ 90~(95)\,\% \ confidence \ level} \, ,}
  \end{array}
\end{equation*}
where the first quoted uncertainty is statistical, the second is systematic, and the last is due to the precision with which the normalisation channel branching fraction is known. The upper limits on the $\Xi_b^0$ include a factor corresponding to the currently unknown ratio of the $\Xi_b^0$ and $\Lambda_b^0$ production fractions.

\subsection{$C\!P$ asymmetry measurements}

The simultaneous extended maximum likelihood fit is modified to allow the determination of the raw asymmetry, defined as
\begin{equation*}
  \label{eq:ACP_master}
  {\cal A}^{\rm raw}_{CP} = \frac{N^{\rm corr}_{f} - N^{\rm corr}_{\bar{f}}}{N^{\rm corr}_{f} + N^{\rm corr}_{\bar{f}}}\,,
\end{equation*}
where $N^{\rm corr}_{f}$ ($N^{\rm corr}_{\bar{f}}$) is the efficiency-corrected yield for $\Lambda_b^0$ ($\bar{\Lambda}_b^{0}$) decays.
The use of the efficiency-corrected yields accounts for the possibility that there may be larger $C\!P$ violation effects in certain regions of phase-space.

To measure the parameter of the underlying $C\!P$ violation, the raw asymmetry has to be corrected for possible small detection (${\cal A}_{\rm D}$) and production (${\cal A}_{\rm P}$) asymmetries,
${\cal A}_{CP} = {\cal A}^{\rm raw}_{CP} - \left({\cal A}_{\rm P} + {\cal A}_{\rm D}\right)$.
This can be conveniently achieved with the $\Lambda_b^0 \rightarrow \left( \Lambda \pi^+ \right)_{\Lambda_c^+} \pi^-$ control mode, which is expected to have negligible $C\!P$ violation.
Since this mode shares the same initial state as the decay of interest, it has the same production asymmetry; moreover, the final-state selection differs only in the PID requirements and therefore most detection asymmetry effects also cancel.

Thus, the results for the phase-space integrated $C\!P$ asymmetries, with correlations taken into account, are
\begin{eqnarray*}
  {\cal A}_{CP}(\Lambda_b^0 \rightarrow \Lambda K^+\pi^-) & = & -0.53 \pm 0.23 \pm 0.11 \, ,\\
  {\cal A}_{CP}(\Lambda_b^0 \rightarrow \Lambda K^+K^-)  & = & -0.28 \pm 0.10 \pm 0.07 \, ,
\end{eqnarray*}
where the uncertainties are statistical and systematic, respectively.

\section{Summary}

The $\Lambda_b^0 \rightarrow \Lambda K^+\pi^-$ and $\Lambda_b^0 \rightarrow \Lambda K^+K^-$ decay modes are observed for the first time, and their branching fractions and $C\!P$ asymmetry parameters are measured.
No evidence is seen for $C\!P$ violation in the phase-space integrated decay rates of these modes.
Evidence is seen for the $\Lambda_b^0 \rightarrow \Lambda \pi^+\pi^-$ decay, and limits are set on the branching fractions of all $\Xi_b^0$ decays under study.

Recent theoretical predictions of $\Lambda_b^0 \rightarrow \Lambda f_0$ decays indicate that these modes could dominate the $\Lambda_b^0 \rightarrow \Lambda \pi^+\pi^-$ decay rate, whereas the $\Lambda_b^0 \rightarrow \Lambda K^+K^-$ decay rate cannot be explained purely by the $\Lambda_b^0 \rightarrow \Lambda \phi$ decay~\cite{Hsiao2016,Geng2016}.

\section*{Acknowledgments}

This work was supported in part by the European Research Council under FP7 and by the United Kingdom's Science and Technology Facilities Council.


\section*{References}

\begin{thebibliography}{1}

\bibitem{Aaltonen:2008hg}
T.~Aaltonen et~al.~(CDF Collaboration)
\newblock {\em Phys. Rev. Lett.}, 103, 031801,\\
\newblock {arXiv:0812.4271,}
\newblock{(2009).}

\bibitem{LHCb-PAPER-2016-002}
R.~Aaij et~al.~(LHCb Collaboration)
\newblock {\em Submitted to PLB,}\\
\newblock {LHCb-PAPER-2016-002, arXiv:1603.02870,}
\newblock {(2016)}.

\bibitem{LHCb-PAPER-2013-061}
R.~Aaij et~al.~(LHCb Collaboration)
\newblock {\em JHEP} 04, 087,
\newblock arxiv:1402.0770,
\newblock{(2014).}

\bibitem{LHCb-PAPER-2014-044}
R.~Aaij et~al. (LHCb Collaboration)
\newblock {\em Phys. Rev.}, D90, 112004,
\newblock{arXiv:1408.5373,}
\newblock {(2014).}

\bibitem{Alves:2008zz}
A.~A. Alves~Jr. et~al.~(LHCb Collaboration)
\newblock {\em JINST}, 3, S08005 (2008).

\bibitem{Aaij:2016nrq}
R.~Aaij et~al.~(LHCb Collaboration)
\newblock {\em JHEP} 05, 081,
\newblock {arXiv:1603.00413,}
\newblock {(2016)}.

\bibitem{Pivk:2005}
M.~Pivk and F.~Le Diberder
\newblock {\em Nucl.Instrum.Meth.}, A555, 356-369,
\newblock arXiv:0402083,
\newblock (2005).

\bibitem{Hsiao2016}
Y.~K.~Hsiao et~al.
\newblock {arXiv:1604.04043,}
\newblock {(2016)}.

\bibitem{Geng2016}
C.~Q.~Geng et~al.
\newblock {arXiv:1603.06682,}
\newblock {(2016)}.

\end{thebibliography}
\end{document}